\documentclass[aps,12pt,final,notitlepage,oneside,onecolumn,%
nobibnotes,nofootinbib,noshowpacs,%
centertags,noshowkeys,amssymb]{revtex4}

\usepackage{graphicx}
\usepackage{amsmath,amsfonts}

\newcommand{\be}{\begin{equation}}
\newcommand{\ee}{\end{equation}}
\newcommand{\ba}{\begin{eqnarray}}
\newcommand{\ea}{\end{eqnarray}}
\newcommand{\nn}{\nonumber\\ }

\newcommand{\vecc}[1]{\mbox{\boldmath $#1$}}

\renewcommand{\Re}{\,\textmd{Re}\,}
\newcommand\m[1]{\mathrm {#1}}
\def\dd{\mathrm d}
\def\ii{\mathrm i}
\DeclareMathOperator{\sign}{sgn}

\begin{document}

\title{Multiple exchanges in lepton pair production in
high--energy heavy ion collisions}

\author{E.~Barto{\v s}}
\altaffiliation{On leave of absence from Dept. of Theoretical
Physics, Comenius University, 84248 Bratislava, Slovakia}%
\affiliation{Joint Institute for Nuclear Research, 141980 Dubna,
Russia}

\author{S.~R.~Gevorkyan}
\altaffiliation{Yerevan Physics Institute, 375036, Yerevan,
Armenia}%
\affiliation{Joint Institute for Nuclear Research, 141980 Dubna,
Russia}

\author{E.~A.~Kuraev}
\affiliation{Joint Institute for Nuclear Research, 141980 Dubna,
Russia}

\author{N.~N.~Nikolaev}
\altaffiliation{ Institute f\"ur Kernphysik, Forshungszentrum
J\"ulich, D-52425 J\"ulich, Germany}%
\affiliation{L.~D.~Landau Institute for Theoretical Physics,
142432 Chernogolovka, Russia}

\begin{abstract}
\begin{center} {\bf Abstract} \end{center}
The recent analysis of nuclear distortions in DIS off nuclei
revealed a breaking of the conventional hard factorization for
multijet observable. The related pQCD analysis of distortion
effects for jet production in nucleus-nucleus collisions is as yet
lacking. As a testing ground for such an analysis we consider the
Abelian problem of higher order Coulomb distortions of the
spectrum of lepton pairs produced in peripheral nuclear
collisions. We report an explicit calculation of the contribution
to the lepton pair production in the collision of two photons from
one nucleus with two photons from the other nucleus, $2\gamma +
2\gamma \to \m{l}^+\m{l}^-$. The dependence of this amplitude on
the transverse momenta has a highly nontrivial form the origin of
which can be traced to the mismatch of the conservation of the
Sudakov components for the momentum of leptons in the Coulomb
field of the oppositely moving nuclei. The result suggests that
the familiar eikonalization of Coulomb distortions breaks down for
the oppositely moving Coulomb centers, which is bad news from the
point of view of extensions to the pQCD treatment of jet
production in nuclear collisions. On the other hand, we notice
that the amplitude for the $2\gamma + 2\gamma \to \m{l}^+\m{l}^-$
process has a logarithmic enhancement for the lepton pairs with
large transverse momentum, which is absent for $n\gamma + m\gamma
\to \m{l}^+\m{l}^-$ processes with $m,n > 2$.

We discuss the general structure of multiple exchanges and show
how to deal with higher order terms which cannot be eikonalized.
\end{abstract}

\maketitle

\section{Introduction}

The exact theory of Coulomb distortions of the spectrum of
ultrarelativistic lepton pairs photoproduced in the Coulomb field
of the nucleus has been developed by Bethe and Maximon \cite{BM}.
It is based on the description of leptons by exact solutions of
the Dirac equation in the Coulomb field (see e.g., the textbook
\cite{LL}). In the Feynman diagram language one has to sum
multiphoton exchanges between produced electrons and positrons and
the target nucleus. For ultrarelativistic leptons this reduces to
the eikonal factors in the impact parameter representation. In the
momentum space the same eikonal form leads to simple recurrence
relations between the $(n+1)$ and $n$--photon exchange amplitudes
\cite{IM}, the incoming photon can be either real or virtual.
There are two fundamental points behind these simple results:
\begin{itemize}
    \item[i)] The lightcone momenta of ultrarelativistic leptons
are conserved in multiple scattering process (i.e., if the nucleus
moves along the $n_-$ lightcone and the produced leptons move
along the $n_+$ lightcone, then the $p_+$ components of the lepton
momenta are conserved).
    \item[ii)] The s--channel helicity of leptons is conserved in
high energy QED (see the textbook \cite{LL}). It is the last
property by which distortions reduce to a simple eikonal factor.
\end{itemize}

The same properties allow one to cast the pair production cross
section in the dipole representation \cite{GG}. They have also
been behind the color dipole pQCD analysis of nuclear distortions
and the derivation of nonlinear $k_{\perp}$--factorization for
multijet hard processes in DIS off nuclei \cite{NSZZ}.

As was shown in \cite{NSZZ2004}, in certain cases of practical
interest the so--called abelianization takes place. Specifically,
the hard dijet production in hadron--nucleus collision is
dominated by a hard collision of an isolated parton from the beam
hadron simultaneously with many gluons from the nucleus which
belong to different nucleons of a target nucleus. None the less,
at least for the single--particle spectra, the interaction with a
large number of nuclear gluons can be reduced to that with a
single gluon from the collective gluon field of a nucleus, i.~e.,
the nonlinear $k_{\perp}$--factorization reduces to the linear one
and in terms of the collective glue one only needs to evaluate the
familiar Born cross sections. The extension of nonlinear
$k_{\perp}$--factorization for hard processes from hadron--nucleus
collisions to collisions of ultrarelativistic nuclei is a
formidable task which has not been properly addressed so far. The
lightcone QED and QCD share many properties, and here we address a
much simpler, Abelian, problem of Coulomb distortions of lepton
pairs produced in peripheral collisions of relativistic nuclei.

The process of lepton pair production in the Coulomb fields of two
colliding ultrarelativistic heavy ions was intensively
investigated in the past years
\cite{BML,SW,ERG,ISS,LM,BGKN1,BGKN2,GK}. Such activity is mainly
connected with new possibilities opened with operation of such
facilities as RHIC and LHC. Despite the high activity in this area
the issue of correct allowance for final state interaction of
produced leptons with the colliding ion Coulomb fields is lacking
yet. The main results obtained so far in this direction are the
following:
\begin{itemize}
    \item[i)] The produced lepton pair interacts
with the Coulomb fields of the ion and the corresponding
corrections have a noticeable impact on the cross section of the
process under consideration at finite energies \cite{ISS}.
    \item[ii)] The perturbation series corresponding to multiple
interaction of a produced pair with Coulomb fields can be summed
and the result can be cast in the eikonal--like form \cite{GK}, if
one restricts ourself to terms growing with energy in the cross
section \cite{BGKN1}. In QED such an approximation can be
considered as satisfactory one, but it does not work in QCD and
the problem of higher order corrections in pair production demands
further investigation.
\end{itemize}

In our paper \cite{BGKN1}, we cited the amplitude $M^{(2)}_{(2)}$
which is irrelevant in leading and next-to leading logarithmic
approximations in QED. Nevertheless, the knowledge of such a kind
contributions becomes important for similar processes in QCD with
multigluon exchanges between the color constituents of each of the
colliding hadrons and the created quark--antiquark pair. Thus, the
main motivation of the present paper is a further investigation of
multiple exchanges and their impact on the lepton pair yield in
the ultrarelativistic heavy ion collisions, an issue which is
useful not only in understanding the electromagnetic processes,
but has a wide application in QCD.

We did not consider the case when one of the ions radiates a
single photon and other one radiates an arbitrary number of
photons absorbed by a created pair \cite{GK}. The photon exchanges
between the ions also were not taken into account \cite{BGKN2}.

Our paper is organized as follows. In Sec.\ref{sec:22}, we
consider the case when each of the colliding ions radiated two
photons which created the lepton pair. We derived the relevant
amplitude $M_{(2)}^{(2)}$ using the powerful Sudakov technique
well suited for calculations of the processes at high energies.

In Sec.\ref{sec:wa}, we studied the wide--angle limit in pair
production kinematics corresponding to the case of large
transverse momenta of pair components. In these limits the results
are much more transparent than in the general case, as can be seen
from the form of the differential cross section which is also
presented.

In Sec.\ref{sec:mp}, we discuss the generalization of the process
under consideration to the case, when the number of exchanged
photons by each ion exceeds two.

\section{The lepton pair production}
\label{sec:22}

We are interested in the process of lepton pair production in the
collision of two relativistic nuclei $A$, $B$ with charge numbers
$Z_1,Z_2$
\begin{equation} \label{eq:proc}
 A(p_1)+B(p_2) \to \m{l}^-(q_-)+\m{l}^+(q_+)+A(p'_1)+B(p'_2),
\end{equation}
with kinematical invariants
\begin{gather}
s=(p_1+p_2)^2,\quad q_1^2=(p_1-p_1')^2,\quad q_2^2=(p_2-p_2')^2\nn
s_1=(q_++q_-)^2,\quad p_1^2=p_1'^2=M_1^2, \quad
p_2^2=p_2'^2=M_2^2, \quad q_\pm^2=m^2.
\end{gather}
We are interested in peripheral kinematics, i.~e.,
\begin{gather}
s\gg M_1^2,\, M_2^2,\, |q_1^2|,\, |q_2^2|, \gg m^2
\end{gather}
which corresponds to small scattering angles of ions $A$ and $B$.

It is convenient to use the Sudakov parameterization for all
4--momenta entering the process (\ref{eq:proc})
\begin{gather}
q_1=a_1\tilde{p}_2+b_1\tilde{p}_1+q_{1\bot},\quad q_2=
a_2\tilde{p}_2+b_2\tilde{p}_1+q_{2\bot},\nn
k_1=\alpha_1\tilde{p}_2+\beta_1\tilde{p}_1+k_{1\bot},\quad
k_2=\alpha_2\tilde{p}_2+\beta_2\tilde{p}_1+k_{2\bot},\nn
q_\pm=\alpha_\pm\tilde{p}_2+\beta_\pm\tilde{p}_1+q_{\pm\bot},
\end{gather}
with lightcone 4--vectors $\tilde{p}_{1,2}$ obeying the conditions
$$\tilde{p}_1^2=\tilde{p}_2^2=0,\quad\tilde{p}_{1,2}. q_\bot=0,\quad
2\tilde{p}_1.\tilde{p}_2=s.$$

\subsection{The pair production by 4--photons}

\begin{figure}[ht]
\begin{center}
\includegraphics[scale=.85]{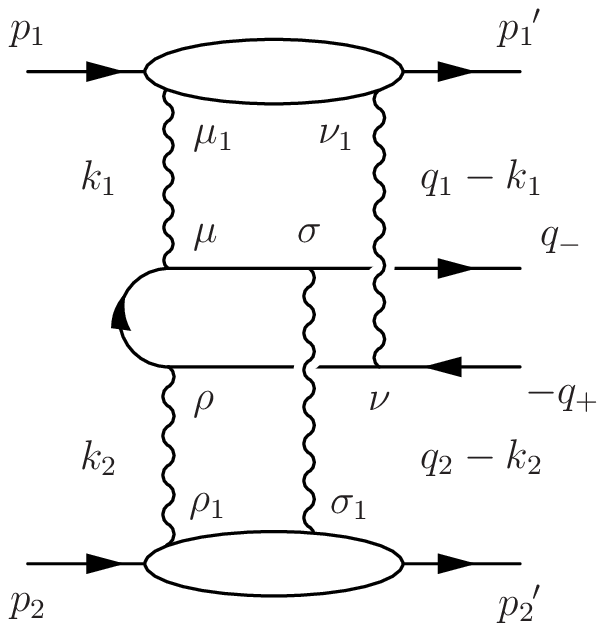}
\end{center}
\caption{\label{fig:1} Typical Feynman diagram for the amplitude
$M_{(2)}^{(2)}$}.
\end{figure}

Let us consider the creation of the lepton pair by four virtual
photons (Fig.~\ref{fig:1}). The photons with momenta $k_1$,
$q_1-k_1$ (in the latter article, referred to as photons $1$ and
$2$) are emitted by the ion A and the photons with momenta $k_2$,
$q_2-k_2$ (referred as the photons $3$ and $4$) by the ion B. The
main contribution to the cross section gives the following regions
of the Sudakov variables:
\begin{gather}
 \alpha_1\ll \beta_1 \sim b_1,\quad \beta_++\beta_-=b_1, \nn
 \beta_2\ll\alpha_2\sim a_2,\quad \alpha_++\alpha_-=a_2, \label{eq:mc}\\
 |a_1|\ll a_2,\quad |b_2|\ll b_1,\quad q_{i\bot}=\vecc{q}_i,\quad
 \vecc{q}_1+\vecc{q}_2=\vecc{q}_++\vecc{q}_-,\nn \nonumber
\alpha_\pm=\frac{\vecc{q}_\pm^2}{s\beta_\pm},\quad \vecc
q_\pm^2\gg m^2.
\end{gather}
Hereinafter $\vecc{q}_i$ denotes the 2--dimensional transverse
part of any considered momenta. For definitness, we suggest
$\beta_+,\:\beta_->0$, which corresponds to the situation when the
pair moves along the ion $A$ (the momentum $p_1$). Bearing in mind
a possible extension to pQCD we neglect the lepton masses whenever
appropriate.

\begin{figure}[ht]
\begin{center}
\includegraphics[scale=.9]{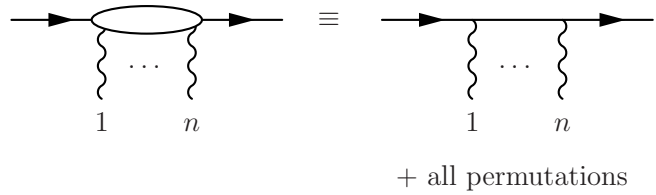}
\end{center}
\caption{\label{fig:2} The notation for the permutations of $n$
virtual photons emitted by the heavy ion.}
\end{figure}

The contribution to the matrix element of such a set of Feynman
diagrams (FD) reads
\begin{multline}
M^{(2)}_{(2)}=\ii s\frac{(Z_1Z_2)^2(4\pi\alpha)^4}{(2\pi)^8}
\int\frac{\dd^4k_1\dd^4k_2}
{k_1^2k_2^2(q_1-k_1)^2(q_2-k_2)^2}\\\label{eq:matrix}
\times\frac{1}{s}\bar{u}^\eta(p_1')O_1^{\mu_1\nu_1}u^\eta(p_1)
\bar{u}^\lambda(p_2') O_2^{\rho_1\sigma_1}u^\lambda(p_2)
\bar{u}(q_-)T^{\mu\nu\rho\sigma}v(q_+) g_{\mu\mu_1}g_{\nu\nu_1}
g_{\rho\rho_1}g_{\sigma\sigma_1}.
\end{multline}
To see the proportionality of the matrix element (\ref{eq:matrix})
to invariant energy $s$, we use the Gribov representation for
virtual photon Green functions
\begin{equation}
g_{\mu\mu_1}g_{\nu\nu_1}g_{\rho\rho_1}g_{\sigma\sigma_1}\approx
\Big(\frac{2}{s}\Big)^4
p_{1\mu}p_{1\nu}p_{1\rho_1}p_{1\sigma_1}p_{2\mu_1}p_{2\nu_1}
p_{2\rho}p_{2\sigma}.
\end{equation}
Numerators of Green functions of the nuclei $A$ can be written as
$s^2N_1$ with $N_1=\bar{u}^\eta(p_1')\hat{p}_2u^\eta(p_1)/s$,
$\sum\limits_\eta|N_1|^2=2$ and a similar expression takes place
for the nuclei $B$. The denominators of virtual photon Green
functions in the considered kinematics depend only on transverse
components of the corresponding 4-vectors, thus
$$
k_1^2k_2^2(q_1-k_1)^2(q_2-k_2)^2=
\vecc{k}_1^2\vecc{k}_2^2(\vecc{q}_1-\vecc{k}_1)^2
(\vecc{q}_2-\vecc{k}_2)^2.
$$
There are 24 FD contributing to $M^{(2)}_{(2)}$. Instead of them
it is convenient to consider $24\ast 2\ast 2=96$ FD which take as
well the permutations of emission and absorption points of
exchanged photons to the nuclei (Fig.~\ref{fig:2}). Then the
result must be divided by $(2!)^2$. This trick
\cite{Gribov:1970ik} provides the convergence of integrals over
$\beta_2$
\begin{equation}
\frac{1}{2\pi \ii}\int\limits_{-\infty}^\infty \dd\beta_2
\bigg[\frac{s}{s\beta_2-c+\ii0}+\frac{s}{-s\beta_2-d+\ii0}\bigg]=-1,
\end{equation}
and a similar integral over the variable $\alpha_1$. After all
operations we can write the matrix element in the form
\begin{equation} \label{eq:dva}
M_{(2)}^{(2)}=\ii s\frac{(16\pi\alpha^2Z_1Z_2)^2N_1N_2}{(2!)^2}
\int\frac{\dd^2\vecc{k}_1\dd^2\vecc{k}_2}{\pi^2}
\frac{\bar{u}(q_-)R
v(q_+)}{\vecc{k}_1^2\vecc{k}_2^2(\vecc{q}_1-\vecc{k}_1)^2
(\vecc{q}_2-\vecc{k}_2)^2},
\end{equation}
with
\begin{equation*}
 R=\frac{1}{s}\int\frac{\dd\beta_1\dd\alpha_2}{(2\pi
\ii)^2}p_{1\mu}p_{1\nu}p_{2\rho}p_{2\sigma} T^{\mu\nu\rho\sigma}.
\end{equation*}

\subsection{The classification of diagrams} \label{classif}

\begin{figure}[ht]
\begin{center}
\includegraphics[scale=.85]{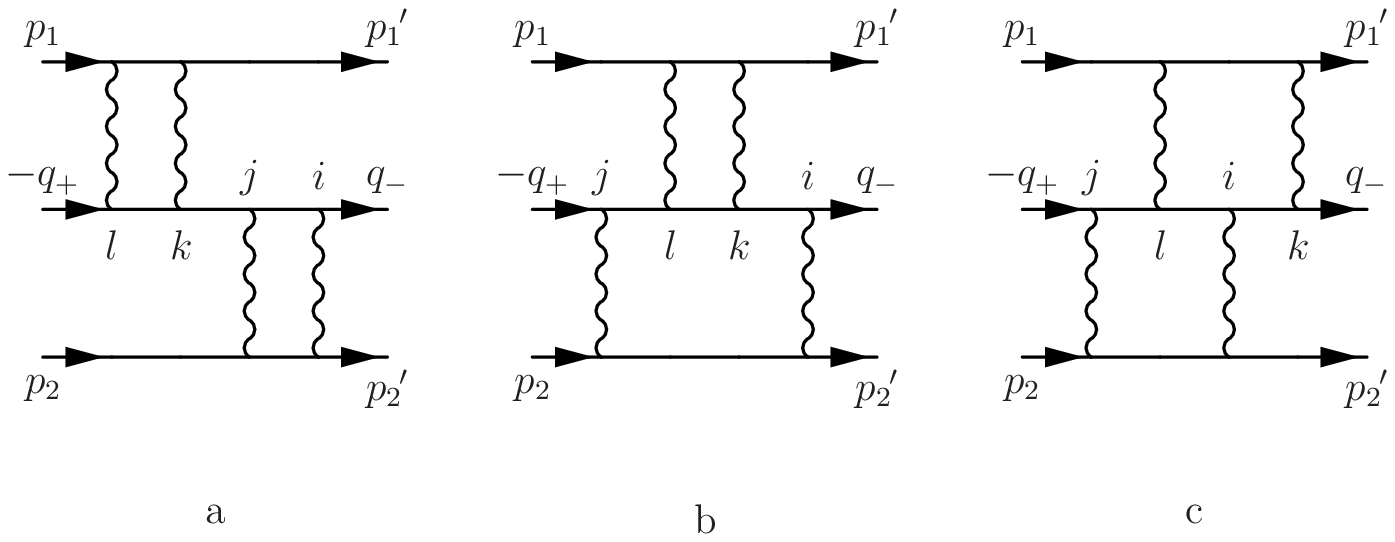}
\end{center}
\caption{\label{fig:3} The set of basic Feynman diagrams for the
amplitude $M_{(2)}^{(2)}$.}
\end{figure}

\begin{figure}[ht]
\begin{center}
\includegraphics[scale=.9]{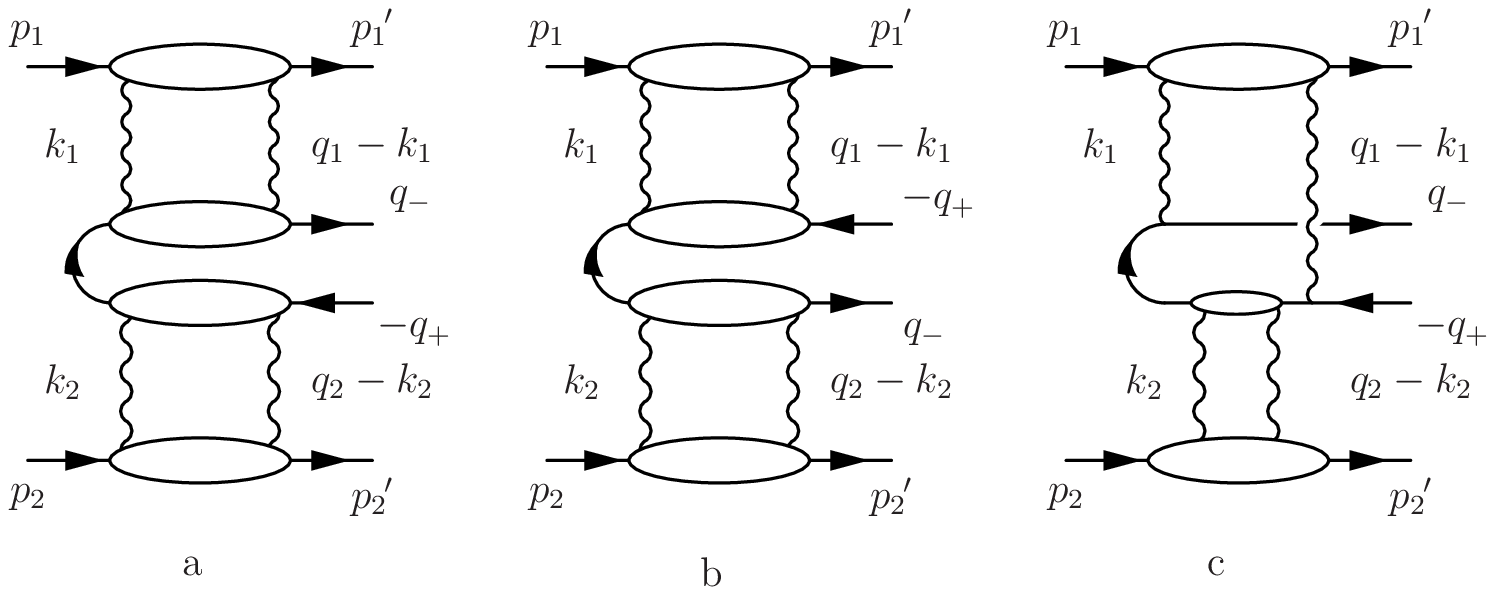} \vspace{8mm}\\
\includegraphics[scale=.9]{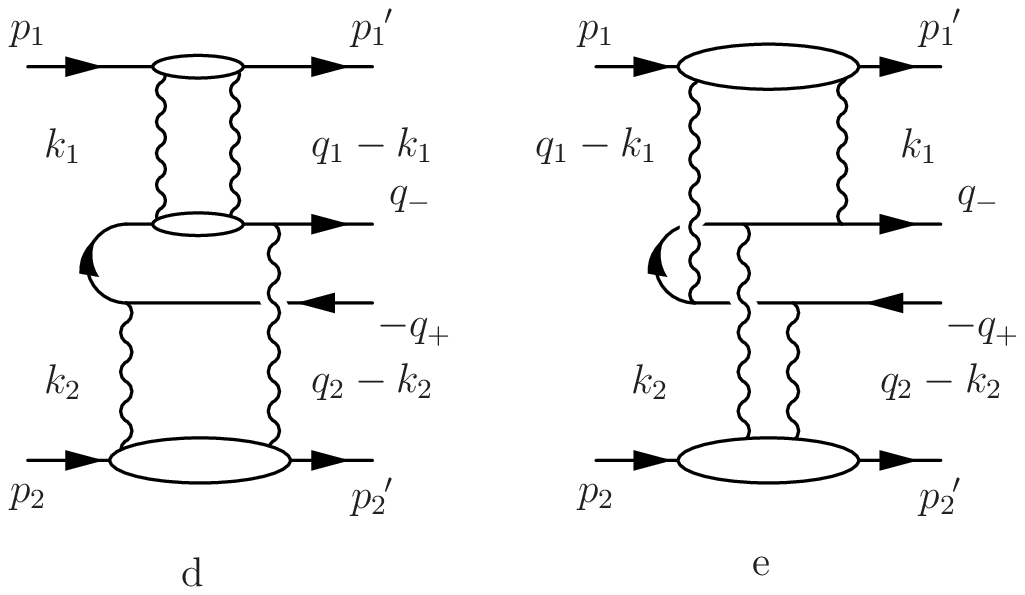}
\end{center}
\caption{\label{fig:4} The Feynman diagrams for the amplitude
$M_{(2)}^{(2)}$.}
\end{figure}

It is convenient to classify FD by order of exchanged photons
absorbed by the lepton world line (Fig.~\ref{fig:3}). We mark them
as $R_{ijkl}$, $R=\sum R_{ijkl}$ with different integers $i,j,k,l$
from one to four, counting from a negative lepton emission point.

a) Consider first the set of 4 FD (Fig.~\ref{fig:4}a) named
$R_{1234}$, $R_{2134}$, $R_{1243}$, $R_{2143}$ in which the
interactions with two nuclei are ordered consecutively against the
lepton line direction. The sum of relevant contributions provides
the convergence of $\beta_1,\alpha_2$ integrations. After a
standard calculation one obtains for this set
\begin{gather}
R_{1234}+R_{2134}+R_{1243}+R_{2143}=
\frac{\beta_-\hat{p}_1(\hat{q}_--
\hat{q}_1)_\bot}{\beta_+\vecc{q}_-^2+
\beta_-(\vecc{q}_--\vecc{q}_1)^2}\frac{\hat{p}_2}{s}
=-B\frac{\hat{p}_2}{s},\nn B=\frac{\hat{q}_{-\bot}(\hat{q}_--
\hat{q}_1)_\bot}{\beta_+\vecc{q}_-^2+
\beta_-(\vecc{q}_--\vecc{q}_1)^2}.   \label{eq:er}
\end{gather}
The last equality in (\ref{eq:er}) is the result of Dirac equation
for massless particles
\begin{equation}
\bar{u}(q_-)\beta_-\hat{p}_1\hat{p}_2=
-\bar{u}(q_-)\hat{q}_{-\bot}\hat{p}_2.
\end{equation}

A similar result as in (\ref{eq:er}) is achieved for the set of
crossing diagrams (Fig.~\ref{fig:4}b) relevant to $R_{3412}$,
$R_{3421}$, $R_{4312}$, $R_{4321}$ terms in the amplitude with
only the replacement $B\to\tilde B$ where for $\tilde B$ stands
\begin{gather}
\tilde B=\frac{(-\hat{q}_++
\hat{q}_1)_\bot\hat{q}_{+\bot}}{\beta_-\vecc{q}_+^2+
\beta_+(\vecc{q}_1-\vecc{q}_+)^2}.   \label{eq:err}
\end{gather}

b) Let us now consider the set of the diagrams $R_{1342}$,
$R_{1432}$, $R_{2341}$, $R_{2431}$(Fig.~\ref{fig:4}c) and
$R_{3124}$, $R_{3214}$, $R_{4123}$, $R_{4213}$
(Fig.~\ref{fig:4}d), where exchanges with the ion B(A) are
attached to the lepton line between the interaction with the ion
A(B).

For definiteness, consider the sum $R_{1342}+R_{1432}$. Using the
relevant denominators of the lepton line one obtains the following
integrals over $\beta_1$, $\alpha_2$:
\begin{align}\label{eq:sign}
&\int\frac{\dd\beta_1}{2\pi \ii}\frac{1}
{s\alpha_-(\beta_--\beta_1)-(\vecc{q}_--\vecc{k}_1)^2+\ii 0}\nn
&\times\frac{1}{-s\alpha_+(\beta_--\beta_1)-
(-\vecc{q}_++\vecc{q}_1-\vecc{k}_1)^2+\ii 0} \nn
&\times\int\frac{\dd\alpha_2}{2\pi
\ii}\Big[\frac{s(\beta_--\beta_1)}
{s(\beta_--\beta_1)(\alpha_--\alpha_2)-(\vecc q_--\vecc k_1)^2
+\ii 0}\nn
&+\frac{s(\beta_--\beta_1)}{s(\beta_--\beta_1)(-\alpha_++\alpha_2)-
(-\vecc q_++\vecc q_1-\vecc k_1)^2+\ii 0}\Big].
\end{align}
The second integral after closing the integration contour in the
lower half plane gives the function $\sign(\beta_--\beta_1)$, thus
(\ref{eq:sign}) becomes
\begin{align}
&\int\frac{\dd\beta_1}{2\pi \ii}\, \frac{\sign(\beta_1-\beta_-)}
{s\alpha_-(\beta_--\beta_1)-(\vecc q_--\vecc k_1)^2 +\ii 0}\nn
&\times \frac{1}{(-s\alpha_+(\beta_--\beta_1)-(-\vecc q_++\vecc
q_1-\vecc k_1)^2+\ii 0)}.
\end{align}
Using the relation
\begin{equation}
\int\limits_{-\infty}^{\infty}\frac{\dd x}{2\pi\ii}
\frac{\sign(x)} {(-ax-b+\ii 0)(cx-d+\ii 0)}=
\frac{1}{\pi\ii(ad+bc)}\ln\frac{ad}{bc},
\end{equation}
we obtained the following result:
\begin{subequations}
\begin{align}
R_{1342}&+R_{1432}+R_{2341}+R_{2431}= \nn \label{eq:i}%
\quad&  \frac{\hat{p}_1}{\ii\pi s} \bigg(\frac{(\hat{q}_--
\hat{k}_1)_\bot (-\hat{q}_++ \hat{q}_1-\hat{k}_1)_\bot}
{\alpha_+(\vecc{q}_--\vecc{k}_1)^2 +
\alpha_-(-\vecc{q}_++\vecc{q}_1-\vecc{k}_1)^2}
\ln\frac{\alpha_+(\vecc{q}_--\vecc{k}_1)^2}
{\alpha_-(-\vecc{q}_++\vecc{q}_1-\vecc{k}_1)^2} \nn \quad+&
\frac{(\hat{q}_-- \hat{q}_1+\hat{k}_1)_\bot (-\hat{q}_++
\hat{k}_1)_\bot} {\alpha_+(\vecc{q}_--\vecc{q}_1+\vecc{k}_1)^2 +
\alpha_-(-\vecc{q}_++\vecc{k}_1)^2}
\ln\frac{\alpha_+(\vecc{q}_--\vecc{q}_1+\vecc{k}_1)^2}
{\alpha_-(-\vecc{q}_++\vecc{k}_1)^2}\bigg),
\\
R_{3124}&+R_{3214}+R_{4123}+R_{4213}= \nn \label{eq:ii}%
\quad&  \frac{\hat{p}_2}{\ii\pi s} \bigg(\frac{(\hat{q}_--
\hat{k}_2)_\bot (-\hat{q}_++ \hat{q}_2-\hat{k}_2)_\bot}
{\beta_+(\vecc{q}_--\vecc{k}_2)^2 +
\beta_-(-\vecc{q}_++\vecc{q}_2-\vecc{k}_2)^2}
\ln\frac{\beta_-(-\vecc{q}_++\vecc{q}_2-\vecc{k}_2)^2}
{\beta_+(\vecc{q}_--\vecc{k}_2)^2} \nn
\quad+& \frac{(\hat{q}_--
\hat{q}_2+\hat{k}_2)_\bot (-\hat{q}_++ \hat{k}_2)_\bot}
{\beta_+(\vecc{q}_--\vecc{q}_2+\vecc{k}_2)^2 +
\beta_-(-\vecc{q}_++\vecc{k}_2)^2}
\ln\frac{\beta_-(-\vecc{q}_++\vecc{k}_2)^2}
{\beta_+(\vecc{q}_--\vecc{q}_2+\vecc{k}_2)^2}\bigg).
\end{align}
\end{subequations}
It is necessary to point out that the obtained expressions
(\ref{eq:i}-\ref{eq:ii}) are pure imaginary and consequently their
interference with the Born term in the cross section is zero.

c) Consider the case of interactions with different nuclei
alternating along the lepton line, for instance, the amplitude
$R_{1324}$ (Fig.~\ref{fig:4}e). After a bit algebra one obtains
for the relevant numerator
\begin{equation}
N_{1324}=s\hat{p}_1\hat{p}_2(\hat{q}_--\hat{k}_1)_\bot
(\hat{q}_--\hat{k}_1-\hat{k}_2)_\bot
(\hat{q}_--\hat{q}_1-\hat{k}_2)_\bot,
\end{equation}
which is very different from the numerators of the Born like
amplitudes. Specifically, it is the term of higher order in the
running transverse momenta $\vecc k_i$.

The relevant denominators read
\begin{align}
\{1\}&=(q_--k_1)^2+i0=s(\beta_--\beta_1)\alpha_--(\vecc
q_--\vecc k_1)^2+i0,\\
\{2\}&=(q_--k_1-k_2)^2+\ii 0=s(\beta_--\beta_1)
(\alpha_--\alpha_2)-(\vecc{q}_--\vecc{k}_1-\vecc{k}_2)^2+\ii 0,\nonumber\\
\{3\}&=(-q_++q_2-k_2)^2+\ii 0=s(-\beta_+)
(\alpha_--\alpha_2)-(-\vecc{q}_++\vecc{q}_2-\vecc{k}_2)^2+\ii
0.\nonumber
\end{align}
The nonvanishing contribution only emerges if the poles are
located in different $\alpha_2$ half--planes, which takes place
only if $\beta_1 < \beta_-$ ($\beta_\pm>0$). Taking the residue at
the pole \{2\} we find
\begin{equation}
\int\frac{s\dd\alpha_2}{2\pi \ii}\frac{1}{\{2\}\{3\}}=
-\frac{\theta(\beta_--\beta_1)}
{(\beta_1-\beta_-)(-\vecc{q}_++\vecc{q}_2-\vecc{k}_2)^2-\beta_+
(\vecc{q}_--\vecc{k}_1-\vecc{k}_2)^2}.
\end{equation}
Further integration over $\beta_1$ can be done using the relation
\begin{equation}
\int\limits_{-\infty}^{\infty}\frac{\dd x}{2\pi\ii}
\frac{\theta(x)} {(ax-b+\ii 0)(cx+d+\ii 0)}=
\frac{-1}{2(ad+bc)}\left(1+\frac{\ii}{\pi}
\ln\frac{ad}{bc}\right),
\end{equation}
wit the result
\begin{gather}
R_{1324}=-\frac{\beta_-N_{1324}}{2sD_{1324}} \left(1 +
\frac{\ii}{\pi}\ln\frac{ad}{bc} \right),\nn \label{eq:denom}
D_{1324}=\beta_-(\vecc{q}_--\vecc{k}_1)^2(-\vecc{q}_++\vecc{q}_2-
\vecc{k}_2)^2+
\beta_+\vecc{q}_-^2(\vecc{q}_--\vecc{k}_1-\vecc{k}_2)^2=ad+bc.
\end{gather}

The highly nonlinear denominator (\ref{eq:denom}) makes the
contribution from the considered case dramatically different from
the Born like amplitude. Technically, the nonlinearity is not
surprising because of the related nonlinearity of the numerator.
The principal difference from the Born like amplitude is that with
the alternating ordering of interactions we have the situation in
which the $p_+$ component of the lightcone momentum is conserved
in the scattering on one ion but is not conserved in the
scattering on the second ion. Depending on the ordering of
interaction vertices and the order of integrations one will
encounter the mismatch of conservation and nonconservation of the
$p_-$ component of the lightcone momentum.

Similar results can be obtained for other contributions of these
types.

d) The final result reads
\begin{equation} \label{eq:dvadva}
M^{(2)}_{(2)}=\frac{\ii s}{(2!)^2}(16\pi\alpha^2Z_1Z_2)^2N_1N_2
\int\frac{\dd^2\vecc{k}_1}{\pi}
\frac{\dd^2\vecc{k}_2}{\pi}\frac{\bar{u}(q_-)R^{(2)}_{(2)}\frac{\hat{p}_2}{s}v(q_+)}
{\vecc{k}_1^2\vecc{k}_2^2(\vecc{q}_1-\vecc{k}_1)^2(\vecc{q}_2-\vecc{k}_2)^2},
\end{equation}
\begin{align} \label{eq:all}
R_{(2)}^{(2)} =&\phantom{+}
\sum_{n=1}^{2}\frac{[\hat{a}_n\hat{b}_n]_\bot}
{\beta_-\vecc{b}_n^2+\beta_+\vecc{a}_n^2}\nn
&-\sum_{n=3}^{10}\frac{[\hat{a}_n\hat{b}_n\hat{c}_n\hat{d}_n]_\bot}
{2[\beta_-\vecc{b}_n^2\vecc{d}_n^2+\beta_+\vecc{a}_n^2\vecc{c}_n^2]}
\bigg(1+\ii\frac{(-1)^{n+1}}{\pi}\ln\frac{\beta_-\vecc{b}_n^2\vecc{d}_n^2}
{\beta_+\vecc{a}_n^2\vecc{c}_n^2}\bigg)\nn
&+\sum_{n=11}^{12}\ii\frac{(-1)^{n+1}}{\pi}\frac{[\hat{a}_n\hat{b}_n]_\bot}
{\beta_-\vecc{b}_n^2+\beta_+\vecc{a}_n^2}
\ln\frac{\beta_-\vecc{b}_n^2}{\beta_+\vecc{a}_n^2}.
\end{align}

\def\h{@{\ }}

\begin{table}[ht]
  \centering
\begin{tabular}{|\h c \h|\h c \h|\h c \h|\h c \h|\h c \h|\h c \h|}
\hline
  n & $R_{ijkl}$ & $a_n$ & $b_n$ & $c_n$ & $d_n$ \\
\hline
  1 & $R_{(12)(34)}$ & $q_-$ & $q_--q_1$ & --- & --- \\
  2 & $R_{(34)(12)}$ & $q_1-q_+$ & $q_+$ & --- & --- \\
  3 & $R_{1324}$     & $q_-$ & $q_--k_1$ & $q_--k_1-k_2$ & $q_--q_1-k_2$ \\
  4 & $R_{1423}$     & $q_-$ & $q_--k_1$ & $q_--q_2+k_2-k_1$ & $-q_++k_2$ \\
  5 & $R_{2314}$     & $q_-$ & $q_--q_1+k_1$ & $q_--q_1+k_1-k_2$ & $-q_++q_2-k_2$ \\
  6 & $R_{2413}$     & $q_-$ & $q_--q_1+k_1$ & $-q_++k_1+k_2$ & $-q_++k_2$ \\
  7 & $R_{4231}$     & $q_--q_2+k_2$ & $-q_++k_1+k_2$ & $-q_++k_1$ & $q_+$ \\
  8 & $R_{3241}$     & $q_--k_2$ & $q_--q_1+k_1-k_2$ & $-q_++k_1$ & $q_+$ \\
  9 & $R_{4132}$     & $q_--q_2+k_2$ & $q_--q_2+k_2-k_1$ & $-q_++q_1-k_1$ & $q_+$ \\
 10 & $R_{3142}$     & $q_--k_2$ & $q_--k_1-k_2$ & $-q_++q_1-k_1$ & $q_+$ \\
 11 & $R_{3(12)4}$   & $q_--k_2$ & $-q_++q_2-k_2$ & --- & --- \\
 12 & $R_{4(12)3}$   & $q_--q_2+k_2$ & $-q_++k_2$ & --- & --- \\
\hline
\end{tabular}
\caption{\label{tab} The coefficients for formula (\ref{eq:all}).
The brackets denote index permutation, e.~g., $(12)\equiv 12+21$.}
\end{table}

To convince the gauge invariance fulfilment we put the explicit
form for the real part of the amplitude
\begin{align*}
\Re R^{(2)}_{(2)}=
&\phantom{+}\frac{[\hat{q}_-(\hat{q}_--\hat{q}_1)]_\bot}{\beta_+\vecc
q_-^2+\beta_-(\vecc q_--\vecc q_1)^2}
+\frac{[(-\hat{q}_++\hat{q}_1)\hat{q}_+]_\bot}{\beta_-\vecc
q_+^2+\beta_+ (\vecc q_+-\vecc q_1)^2}\\ \displaybreak
&\phantom{+}
\frac{[\hat{q}_-(\hat{q}_--\hat{k}_1)(\hat{q}_--\hat{k}_1-\hat{k}_2)
(\hat{q}_--\hat{q}_1-\hat{k}_2)]_\bot} {2[\beta_-(\vecc q_--\vecc
k_1)^2(-\vecc q_++\vecc q_2-\vecc k_2)^2+ \beta_+\vecc q_-^2(\vecc
q_--\vecc k_1-\vecc k_2)^2]}\\ &-
\frac{[\hat{q}_-(\hat{q}_--\hat{k}_1)(\hat{q}_--\hat{q}_2+\hat{k}_2-
\hat{k}_1)(-\hat{q}_++\hat{k}_2)]_\bot}{2[\beta_-(\vecc q_--\vecc
k_1)^2(-\vecc q_++\vecc k_2)^2+ \beta_+\vecc q_-^2(\vecc q_--\vecc
q_2+\vecc k_2-\vecc k_1)^2]}\\ &-
\frac{[\hat{q}_-(\hat{q}_--\hat{q}_1+\hat{k}_1)(\hat{q}_--\hat{q}_1
+\hat{k}_1-\hat{k}_2)(-\hat{q}_++\hat{q}_2-\hat{k}_2)]_\bot}
{2[\beta_-(\vecc q_--\vecc q_1+\vecc k_1)^2(-\vecc q_++\vecc
q_2-\vecc k_2)^2+ \beta_+\vecc q_-^2(\vecc q_--\vecc q_1+\vecc
k_1-\vecc k_2)^2]}\displaybreak[0]\\ &-
\frac{[\hat{q}_-(\hat{q}_--\hat{q}_1+\hat{k}_1)
(-\hat{q}_++\hat{k}_1+\hat{k}_2)(-\hat{q}_++\hat{k}_2)]_\bot}
{2[\beta_-(\vecc q_--\vecc q_1+\vecc k_1)^2(-\vecc q_++\vecc
k_2)^2+ \beta_+\vecc q_-^2(-\vecc q_++\vecc k_1+\vecc k_2)^2]}\\
&-
\frac{[(\hat{q}_--\hat{q}_2+\hat{k}_2)(-\hat{q}_++\hat{k}_1+\hat{k}_2)
(-\hat{q}_++\hat{k}_1)\hat{q}_+]_\bot}{2[\beta_-\vecc q_+^2(-\vecc
q_++\vecc k_1+\vecc k_2)^2+ \beta_+(-\vecc q_++\vecc k_1)^2(\vecc
q_--\vecc q_2+\vecc k_2)^2]}\\ &-
\frac{[(\hat{q}_--\hat{k}_2)(\hat{q}_--\hat{q}_1+\hat{k}_1-\hat{k}_2)
(-\hat{q}_++\hat{k}_1)\hat{q}_+]_\bot}{2[\beta_-\vecc q_+^2(\vecc
q_--\vecc q_1+\vecc k_1-\vecc k_2)^2+ \beta_+(-\vecc q_++\vecc
k_1)^2(\vecc q_--\vecc k_2)^2]}\displaybreak[0]\\ &-
\frac{[(\hat{q}_--\hat{q}_2+\hat{k}_2)(\hat{q}_--\hat{q}_2+\hat{k}_2
-\hat{k}_1)(-\hat{q}_++\hat{q}_1-\hat{k}_1)\hat{q}_+]_\bot}
{2[\beta_-\vecc q_+^2(\vecc q_--\vecc q_2+\vecc k_2-\vecc k_1)^2+
\beta_+(\vecc q_--\vecc q_2+\vecc k_2)^2(-\vecc q_++\vecc
q_1-\vecc k_1)^2]}\displaybreak[0]\\ &-
\frac{[(\hat{q}_--\hat{k}_2)(\hat{q}_--\hat{k}_1-\hat{k}_2)
(-\hat{q}_++\hat{q}_1-\hat{k}_1)\hat{q}_+]_\bot} {2[\beta_-\vecc
q_+^2(\vecc q_--\vecc k_1-\vecc k_2)^2+ \beta_+(-\vecc q_++\vecc
q_1-\vecc k_1)^2(\vecc q_--\vecc k_2)^2]}.
\end{align*}
Then one can verify that the following condition is satisfied:
\begin{equation}
\Re R^{(2)}_{(2)}=0 \quad \mathrm{if} \quad\vecc{k}_1=0\quad
\mathrm{or}\quad \vecc{k}_2=0\quad \mathrm{or}\quad
\vecc{k}_1=\vecc{q}_1\quad\mathrm{or}\quad \vecc{k}_2=\vecc{q}_2.
\end{equation}
This fact is correct also for the whole amplitude (\ref{eq:all}).
As one can see, this property is crucial for the infrared
convergence in integrations over $\dd^2\vecc{k}_i$ .

Under the loop integration one can make the shift of the
integration variable $\vecc{k}_i\to\vecc{q}_i-\vecc{k}_i$. Then
expression (\ref{eq:all}) for $\Re R^{(2)}_{(2)}$ can be
simplified to
\begin{align} \label{eq:uprost}
\Re R^{(2)}_{(2)}=&\phantom{+}\frac{\hat{q}_{-\bot}(\hat{q}_--
\hat{q}_1)_\bot}{\beta_+\vecc{q}_-^2+
\beta_-(\vecc{q}_--\vecc{q}_1)^2} +\frac{(-\hat{q}_++
\hat{q}_1)_\bot\hat{q}_{+\bot}}{\beta_-\vecc{q}_+^2+
\beta_+(\vecc{q}_1-\vecc{q}_+)^2}\nn &-
2\frac{[\hat{q}_-(\hat{q}_--\hat{k}_1)
(\hat{q}_--\hat{k}_1-\hat{k}_2)(\hat{q}_--\hat{q}_1-\hat{k}_2)]_\bot}
{\beta_-(\vecc q_--\vecc k_1)^2(-\vecc q_++\vecc q_2-\vecc k_2)^2+
\beta_+\vecc q_-^2(\vecc q_--\vecc k_1-\vecc k_2)^2}\nn &-
2\frac{[(-\hat{q}_++\hat{q}_1+\hat{k}_2)
(-\hat{q}_++\hat{k}_1+\hat{k}_2)(-\hat{q}_++\hat{k}_1)\hat{q}_+]_\bot}
{\beta_-\vecc q_+^2(-\vecc q_++\vecc k_1+\vecc k_2)^2+
\beta_+(-\vecc q_++\vecc k_1)^2(\vecc q_--\vecc q_2+\vecc k_2)^2}.
\end{align}
Despite the gauge invariance property is not seen clearly her,e as
in the previous case, the final results after integration over
$k_i$ coincide.

\section{The wide angle limit of the $M^{(2)}_{(2)}$ amplitude}
\label{sec:wa}

Let us consider the behavior of this expression in the case when
the transverse component of lepton momenta is large compared to
the momenta transferred to the ions
\begin{equation}
\vecc{q}_-\approx -\vecc{q}_+=\vecc{q},\quad
|\vecc{q}|\gg|\vecc{q}_{1,2}|.
\end{equation}
In this case, the main contribution to the matrix element gives
the region
\begin{equation}
|\vecc{q}_i|\ll|\vecc{k}_i|\ll|\vecc{q}|.
\end{equation}

The amplitude $M^{(1)}_{(1)}$ reads
\begin{gather}
M^{(1)}_{(1)}=-\ii s
\frac{(8\pi\alpha)^2N_1N_2Z_1Z_2}{\vecc{q}_1^2\vecc{q}_2^2}
\bar{u}(q_-)\frac{R^{(1)}_{(1)}}{s}v(q_+), \\ \nonumber
R^{(1)}_{(1)}=\frac{1}{s}\hat{p}_1\frac{\hat{q}_--\hat{q}_1}
{(q_--q_1)^2}\hat{p}_2+ \hat{p}_2\frac{\hat{q}_1-\hat{q}_+}
{(q_1-q_+)^2}\hat{p}_1=(B-\tilde B)\hat{p}_2.
\end{gather}
For wide angle kinematics one has
\begin{gather}
\frac{1}{s}R^{(1)}_{(1)}=\frac{\hat{p}_2}{s}
\frac{1}{b_1^2(\vecc{q}^2)^2}[2\vecc{q}.\vecc{q}_2[
b_1\hat{q}\hat{q}_1+2\beta_-\vecc{q}.\vecc{q}_1]+\vecc{q}^2
[b_1\hat{q}_1\hat{q}_2+2\beta_+\vecc{q}_1.\vecc{q}_2]],
\end{gather}
with $b_1=\beta_-+\beta_+$,
$\vecc{q}=\vecc{q}_-\approx-\vecc{q}_+$ and $\vecc{q}_{1,2}$ are
the transferred to ions momenta.

For matrix element $M^{(1)}_{(2)}$ we have (in agreement with the
result obtained in paper \cite{GPS})
\begin{gather}
M^{(1)}_{(2)}=-s\frac{2^7\pi^2\alpha^3Z_1Z_2^2N_1N_2}
{\vecc{q}_1^2}\int \frac{d^2\vecc{k}_2}{\pi}\frac{1}{\vecc{k}_2^2
(\vecc{q}_2-\vecc{k}_2)^2}\bar{u}(q_-)R_{(2)}^{(1)}
\frac{\hat{p}_2}{s}v(q_+),
\end{gather}
with
\begin{gather}
R_{(2)}^{(1)}=B+\tilde B-
\frac{(\hat{q}_--\hat{k}_2)_\bot(\hat{q}_--\hat{q}_1-\hat{k}_2)_\bot}
{\beta_-(\vecc{q}_--\vecc{q}_1-\vecc{k}_2)^2+\beta_+(\vecc{q}_--\vecc{k}_2)^2}-
\frac{(\hat{q}_+-\hat{k}_2-\hat{q}_1)_\bot(\hat{q}_+-\hat{k}_2)_\bot}
{\beta_+(\vecc{q}_--\vecc{q}_2+\vecc{k}_2)^2+\beta_-(\vecc{q}_+-\vecc{k}_2)^2}.
\end{gather}
In the considered limit this expression has the form
\begin{align}
R_{(2)}^{(1)}\sim&\frac{1}{b_1\vecc{q}^2}
\Big[(2\beta_-\vecc{q}_-.\vecc{q}_1 +\hat{q}_-\hat{q}_1)
\Big(\frac{4(\vecc{q}_-.\vecc{k}_2)^2}{(\vecc{q}^2)^2}
-\frac{\vecc{k}_2^2}{\vecc{q}^2}\Big)\nn &-
\frac{2\vecc{q}_-.\vecc{k}_2}{\vecc{q}^2}(\hat{k}_2\hat{q}_1+
2\beta_-\vecc{k}_2.\vecc{q}_1)]+ (\beta_-\to\beta_+),\quad
|\vecc{k}_2|\gg|\vecc{q}_2|.
\end{align}
This expression turns to zero after the angular averaging. It can
be shown that the quantity $M^{(1)}_{(3)}$ as well turns to zero
in the limit of wide angles pair production and is proportional to
$|\vecc{q}_2|/|\vecc{q}|\ll 1$, which is in agreement with
\cite{IM}.

\begin{figure}
\begin{center}
\includegraphics[scale=.9]{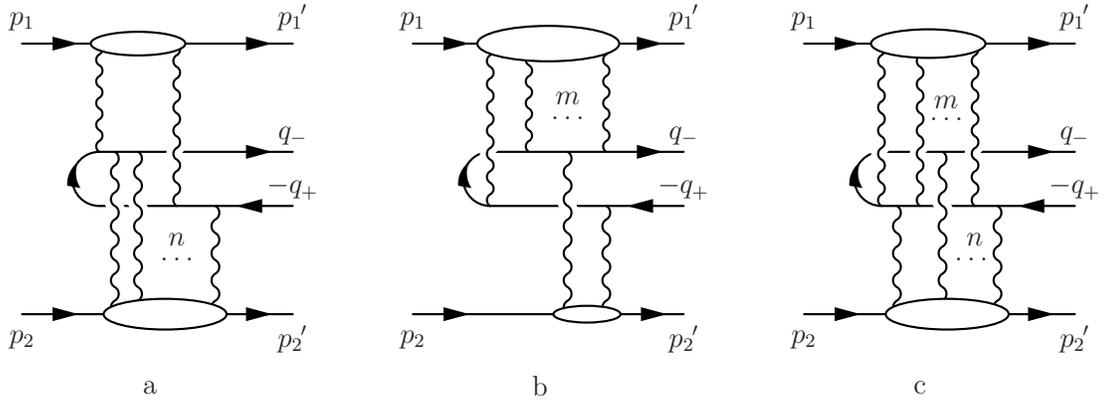}
\end{center}
\caption{\label{fig:5} Some Feynman diagrams for the amplitudes of
the type $M^{(2)}_{(n)}$ (a), $M^{(n)}_{(2)}$ (b) and
$M^{(m)}_{(n)}$ with $m, n\ge 2$.}
\end{figure}

For the considered above amplitude $M_{(2)}^{(2)}$
(\ref{eq:dvadva}) the quantity $R^{(2)}_{(2)}$ plays a role of
cut-off in the region $|\vecc{k}_i|>|\vecc{q}|$. From very general
arguments it can be cast in the form
\begin{equation} \label{eq:tenzor}
\Re R^{(2)}_{(2)}\approx\frac{[k_1^\mu(q_1-k_1)^\nu
k_2^\alpha(q_2-k_2)^\beta]_\bot}{(\vecc
q^2)^2}R_{\mu\nu\alpha\beta},
\end{equation}
with some dimensionless tensor matrix $R_{\mu\nu\alpha\beta}$
independent of $\vecc k_i,\vecc q_i$. Expanding the expression
(\ref{eq:uprost}) one gets
\begin{equation}
\int\frac{\dd^2\vecc{k}_1\dd^2\vecc{k}_2}{\pi^2}\frac{\Re
R^{(2)}_{(2)}} {\vecc{k}_1^2\vecc{k}_2^2(\vecc{q}_1-\vecc{k}_1)^2
(\vecc{q}_2-\vecc{k}_2)^2}\approx
\frac{I}{(\vecc{q}^2)^2}\frac{4(\beta_+-\beta_-)}
{(\beta_-+\beta_+)^2}\ln\frac{\vecc{q}_{max}^2}
{\vecc{q}_1^2}\ln\frac{\vecc{q}_{max}^2}{\vecc{q}_2^2},
\end{equation}
where $I$ is the unit matrix and $q_{max}$ is the upper
integration limit $q_{max}\simeq 1/R$, $R$ is the nucleus radius.
Such enhancement is absent if the number of exchanged photons from
every ion exceeds two (Fig.~\ref{fig:5}). Really, the amplitudes
$M^{(2)}_{(n)}$, $M^{(n)}_{(2)}$, $n>2$ contain only the first
power of large logarithm, whereas $M^{(m)}_{(n)}$, $m,n>2$ do not
contain such a factor at all, because the corresponding loop
momenta integrals are convergent in both infrared and ultraviolet
regions and one can safely put $|\vecc{q}_{1(2)}|=0$ over loop
integrations.

Thus, the differential cross section for the considered kinematics
is determined by the interference term
$(M^{(1)}_{(1)})^*M^{(2)}_{(2)}$ which has the form (for
comparison we present also the Born term)
\begin{align}
\frac{\dd\sigma_{0}}{\dd b_1\dd x}&= \frac{16
(Z_1Z_2\alpha^2)^2}{\pi^4}\;
\frac{(x^2+(1-x)^2)}{\vecc{q}_1^2\vecc{q}_2^2(\vecc{q}^2)^2b_1}\:
\dd^2q_1\dd^2q_2\dd^2q,\\
\label{eq:kve}
\frac{\dd\sigma_{int}}{\dd b_1\dd x}&= \frac{16
(Z_1Z_2\alpha^2)^3}{\vecc{q}_1^2\vecc{q}_2^2\vecc{q}_+^2\vecc{q}_-^2}
\frac{(1-2x)}{b_1}\ln\frac{\vecc{q}_{max}^2}{\vecc{q}_1^2}
\ln\frac{\vecc{q}_{max}^2}{\vecc{q}_2^2}\:
Q\:\dd^2q_1\dd^2q_2\dd^2q_-,\\
&\; Q=\frac{\vecc{q}_-.(\vecc{q}_1-\vecc{q}_-)}{(1-x)\vecc{q}_-^2+
x(\vecc{q}_--\vecc{q}_1)^2}+
\frac{\vecc{q}_+.(\vecc{q}_+-\vecc{q}_1)}{x\vecc{q}_+^2+
(1-x)(\vecc{q}_1-\vecc{q}_+)^2}, \nn \nonumber
&\; x=\frac{\beta_-}{b_1},\quad \epsilon <x,\quad b_1 <
1-\epsilon,\quad \epsilon=\frac{4m^2x(1-x)}{\vecc{q}_\pm^2}.
\end{align}
We note that expression (\ref{eq:kve}) is symmetric under
simultaneous substitutions $q_+\leftrightarrow q_-$ and
$\beta_+\leftrightarrow \beta_-$ due to the C-even character of
the interference.

Finally, from very straightforward generalization of
(\ref{eq:tenzor}) it can be shown that the set of amplitudes with
an odd number of exchanges with one or both nuclei is suppressed
in the limit of wide angle production
\begin{equation} \label{eq:m}
M^{(2m)}_{(2n+1)}\sim O\Big(\frac{\mid\vecc q_1\mid}{\mid\vecc
q\mid}\Big),\quad M^{(2m+1)}_{(2n)}\sim O\Big(\frac{\mid\vecc
q_2\mid}{\mid\vecc q\mid}\Big),\quad M^{(2m+1)}_{(2n+1)}\sim
O\Big(\frac{\mid\vecc q_1\mid\mid\vecc q_2\mid}{\mid\vecc
q^2\mid}\Big).
\end{equation}

\section{Multiphoton exchange}
\label{sec:mp}

\begin{figure}[ht]
\begin{center}
\includegraphics[scale=1]{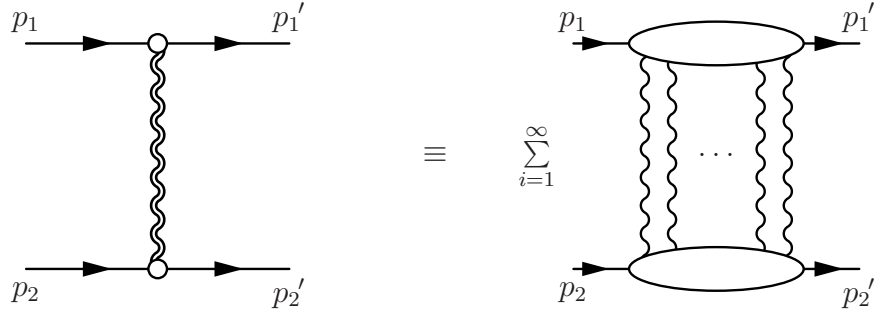}
\end{center}
\caption{\label{fig:6} The representation of all eikonal
exchanges.}
\end{figure}

Let us generalize the above picture for the case of multiple
photon exchanges ($m, n >2$). Using the relation
\begin{gather}
I_n=\frac{1}{\pi^{n-1}}\int\frac{\dd^2\vecc{k}_1\dots\dd^2\vecc{k}_{n-1}}
{(\vecc{k}_1^2+\lambda^2)\dots(\vecc{k}_{n-1}^2+\lambda^2)
((\vecc{q}-\vecc{k}_1- \dots-\vecc{k}_{n-1})^2+\lambda^2)}=
\frac{n\ln^{n-1}(\frac{\vecc q^2}{\lambda^2})}{\vecc{q}^2},
\end{gather}
and taking into account the combinatorial factor $\frac{1}{n!}$
coming from the symmetric integration over $\alpha_i$, $\beta_i$,
one has to replace any single photon exchange by an infinite set
of photons, multiplying the amplitude by the factors of type
$\exp\{\ii\varphi_{i}(\vecc q^2)\}$ with the phase
$\varphi_{i}(\vecc q^2)=\pm\alpha Z_i\ln\frac{\vecc
q^2}{\lambda}$. The scattering of electron and positron differs
only by sign of the phase (positive for electrons) \cite{ERG}.
This replacement is depicted in Fig.~\ref{fig:6} where the double
photon line corresponds to the infinite set of photons.

\begin{figure}[ht]
\begin{center}
\includegraphics[scale=.8]{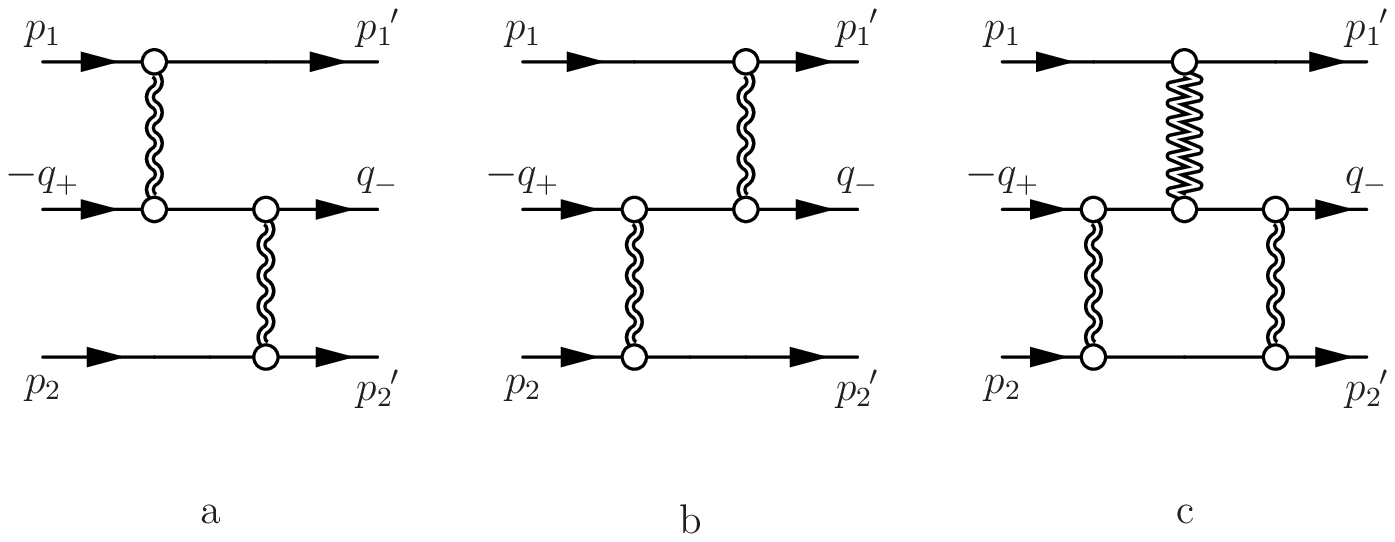} \vspace{8mm}\\
\includegraphics[scale=.8]{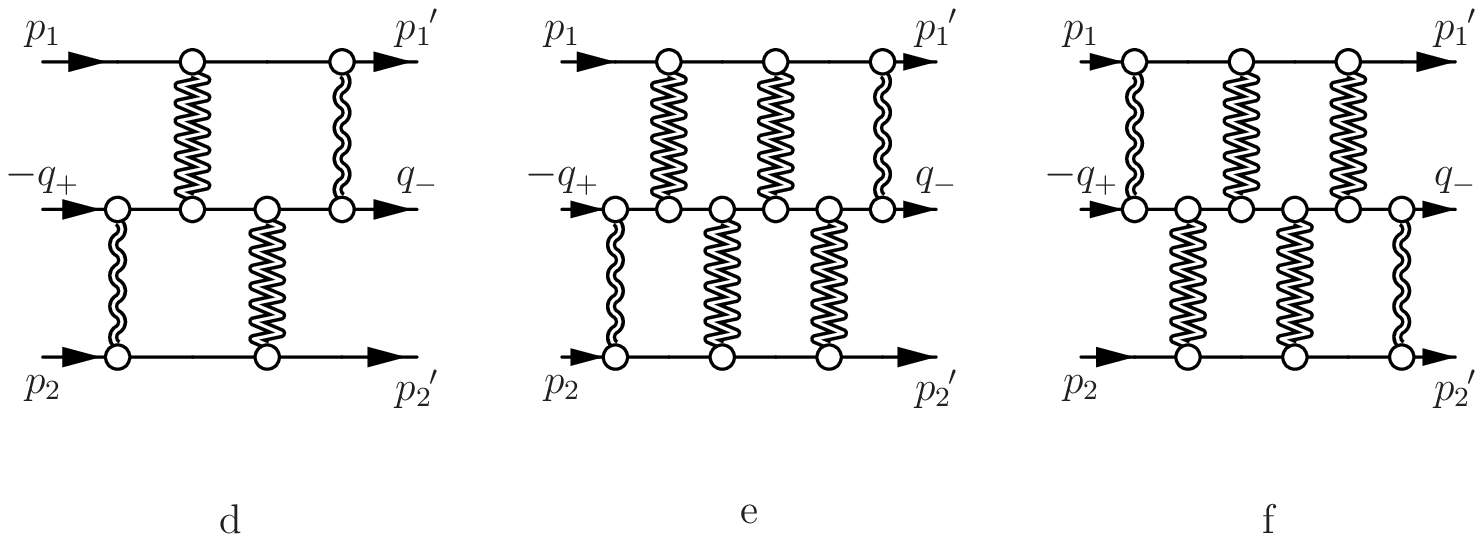}
\end{center}
\caption{\label{fig:7} The Feynman diagrams for the amplitudes
with many photon exchanges. The double photon line represents any
number of exchanged photons, the double zigzag line represents
only the odd number of exchanged photons.}
\end{figure}

Using the same technique as in \cite{BGK} one can see that the
amplitude relevant to Fig.~\ref{fig:7}a and Fig.~\ref{fig:7}b can
be cast in the form
\begin{gather}
\tilde{R}_{(1)}^{(1)}=B\m{e}^{\displaystyle -\ii[\varphi_1(\vecc
q_1^2)-\varphi_2(\vecc q_2^2)]} +\tilde B\m{e}^{\displaystyle
\ii[\varphi_1(\vecc q_1^2)-\varphi_2(\vecc q_2^2)]}.
\end{gather}
The interaction of the electron and the positron with Coulomb
field differs only by signs. Though this expression is infrared
unstable in the case $Z_1\ne Z_2$ the regularization parameter
$\lambda$ enters it in a standard way.

Let us now consider the class of diagrams depicted on
Fig.~\ref{fig:7}c. In subsection \ref{classif}, we obtained the
expressions (\ref{eq:i}, \ref{eq:ii}) for the case $m=n=2$ such
that $\Re R_{1(34)2}=0$. It can be shown that the terms of higher
order with any even number of photons from same nuclei attached to
the lepton world line between two photons from other nuclei do not
contribute to the amplitude of the process under consideration. It
is the consequence of the relation
$(\sign(\alpha))^{2k+1}=\sign(\alpha)$.

The general structure of the amplitude corresponding to
Fig.~\ref{fig:7}c can be constructed using the lowest order
truncated amplitude (without single photon propagators)
$R_{(2)}^{(1)}$
\begin{gather}
\tilde R_{(2)}^{(1)}=\frac{\cos \varphi_1(\vecc q_1^2)}{q_1^2}
\bar{R}_{(2)}^{(1)} \m{e}^{\displaystyle \ii[\varphi_2(\vecc
k^2)-\varphi_2((\vecc q_2-\vecc k)^2)]},\nn%
\bar{R}_{(2)}^{(1)}=\frac{1}{\ii\pi}
\frac{(\hat{q}_--\hat{q}_2+\hat{k})_\bot(-\hat{q}_++\hat{k})_\bot}
{\beta_-(\vecc{q}_+-\vecc{k})^2+\beta_+(\vecc{q}_--\vecc{q}_2+
\vecc{k})^2}\ln\frac{\beta_+(\vecc{q}_--\vecc{q}_2+
\vecc{k})^2}{\beta_-(\vecc{q}_+-\vecc{k})^2}
\end{gather}

The further generalization is obvious. For instance, we cite the
expression corresponding to the diagram depicted on
Fig.~\ref{fig:7}d
\begin{equation}
\tilde R_{(2)}^{(2)}=
\cos\varphi_1(\vecc k_1^2)\, \m{e}^{\displaystyle
-\ii\varphi_1((\vecc q_1-\vecc k_1)^2)}\, \cos\varphi_2(\vecc
k_2^2)\, \m{e}^{\displaystyle  \ii\varphi_2((\vecc q_2-\vecc
k_2)^2)}\, R_{1324}.
\end{equation}

From the above consideration we conclude that the general
structure of the matrix element $M^{(m)}_{(n)}$, corresponding to
$m$ photon exchanges from one ion and $n$ exchanges from other
one, schematically reads
\begin{multline}
M^{(m)}_{(n)}=\ii sN_1N_2(Z_1\alpha)^m(Z_2\alpha)^n
\,\frac{\pi^2}{16n!m!}\\ \times
\int\frac{\dd^2k_1}{\pi}\cdots\frac{\dd^2k_{m-1}}{\pi}
\,\frac{\dd^2\kappa_1}{\pi}\cdots\frac{\dd^2\kappa_{n-1}}{\pi}\,
\frac{1}{\vecc{k}^2_1\dots\vecc{k}^2_m}\,
\frac{1}{\vecc{\kappa}_1^2\dots\vecc\kappa_n ^2}
\bar{u}(q_-)\bar{R}^{(m)}_{(n)}\frac{\hat{p}_2}{s}\,v(q_+),
\end{multline}
where $m$ and $n$ obey the condition $|m-n|\le 1$. At this stage,
we omitted phase factors in the structure ${R}^{(m)}_{(n)}$ (for
clearly understanding the problem), so it can be written in the
form
\begin{gather} \label{eq:long}
\bar{R}^{(m)}_{(n)}=\bar{R}^{(1)}_{(1)}+
\bar{R}^{(1)}_{(2)}+\bar{R}^{(2)}_{(1)}+
\bar{R}^{(2)}_{(2)}+\bar{R}^{(2)}_{(3)}+
\bar{R}^{(3)}_{(2)}+\bar{R}^{(3)R}_{(3)}+
\bar{R}^{(3)L}_{(3)}\dots
\end{gather}
\begin{align}
\bar{R}^{(2)}_{(1)}&=\frac{1}{\ii\pi}
\frac{(\hat{q}_--\hat{k})_\bot(-\hat{q}_++\hat{q}_1-\hat{k})_\bot}
{\alpha_-(-\vecc{q}_++\vecc{q}_1-\vecc{k})^2+\alpha_+(\vecc{q}_--
\vecc{k})^2}\ln\frac{\alpha_+(\vecc{q}_--
\vecc{k})^2}{\alpha_-(-\vecc{q}_++\vecc{q}_1-\vecc{k})^2},\nn%
\bar{R}^{(2)}_{(3)}&=\bar{R}^{(3)}_{(2)}=0,\quad\nn
\bar{R}^{(3)R}_{(3)}&=\frac{1}{c_1+c_2}\Big[3\zeta_2+\frac{1}{2}\ln^2
\frac{c_1}{c_2}\Big],\quad \zeta_2=\frac{\pi^2}{6},\nn%
\qquad &c_1=\beta_-(\vecc q_--\vecc k_1)^2(\vecc q_--\vecc
k_1-\vecc k_2-\vecc \kappa_1)^2(-\vecc q_++\vecc q_2-\vecc
\kappa_1-\vecc \kappa_2)^2,\nn%
\qquad &c_2=\beta_+\vecc q_-^2(\vecc q_--\vecc k_1-\vecc
\kappa_1)^2(\vecc q_--\vecc k_1-\vecc k_2-\vecc \kappa_1-\vecc
\kappa_2)^2,\nn%
\bar{R}^{(3)}_{(4)}&=\frac{1}{d_1+d_2}\Big[3\zeta_2+\frac{1}{2}\ln^2
\frac{d_1}{d_2}\Big]\nn%
\qquad &d_1=\beta_+(\vecc q_--\vecc \kappa_1)^2 (\vecc q_--\vecc
\kappa_1-\vecc \kappa_2-\vecc k_1)^2 (\vecc q_--\vecc
\kappa_1-\vecc \kappa_2-\vecc \kappa_3-\vecc k_1-\vecc k_2)^2,\nn
\qquad &d_2=\beta_-(\vecc q_--\vecc \kappa_1-\vecc k_1)^2 (\vecc
q_--\vecc \kappa_1-\vecc \kappa_2-\vecc k_1-\vecc k_2)^2 (-\vecc
q_++\vecc q_2-\vecc \kappa_1-\vecc \kappa_2-\vecc \kappa_3)^2.
\end{align}
Here $\bar{R}^{(2)}_{(2)}$ is only the second term in the
right--hand side in (\ref{eq:all}) and the index $R(L)$ denotes
two possible configurations of photons for $\bar{R}^{(3)R}_{(3)}$
(Fig.~\ref{fig:7}e) and $\bar{R}^{(3)L}_{(3)}$
(Fig.~\ref{fig:7}f).

In such a way, the general algorithm for construction of an
arbitrary term is transparent. Unfortunately, we cannot obtain the
compact expression for the whole amplitude. The reason is the
increasing nonlinearity of the propagators with the order of
interaction. The behavior of the above denominators is very
different from the Born--like case, where the simplicity of
propagators allows one to obtain eikonal--like expressions.

\section{Conclusions}
The wide angle lepton pair production in peripheral interactions
of ultrarelativistic heavy ions is an archetype reaction for hard
processes in central hadronic hard collisions of heavy nuclei. In
the electromagnetic case, the expansion parameter $Z_{1,2}\,\alpha
\sim 1$ makes the multiple photon collisions, $m\gamma+n\gamma \to
\m{l}^+\m{l}^-$ potentially important ones, likewise the effect of
multiple gluon collisions in central collisions is enhanced by a
large number of nucleons at the same impact parameter. The crucial
issue is whether such multiple photon collisions can be described
by the Born cross section in terms of the collective photon fields
of colliding nuclei or not. We obtained the expression for the
amplitude $2\gamma+2\gamma \to \m{l}^+\m{l}^-$ and show that its
contribution is dominant in a wide angle limit. Our principal
finding is that the amplitude is manifestly of non--Born nature,
which is suggestive of the complete failure of linear
$k_{\perp}$--factorization even in the Abelian case.

We have shown that the terms in perturbation series of the
amplitude for the process of lepton pair production in the Coulomb
fields of two relativistic nuclei relevant to the closed two
photon loops are logarithmically enhanced in this case, while in
higher order terms such enhancement is absent. We presented the
algorithm which allows one to construct the full amplitude in all
orders. The obtained results can be useful in application to the
QCD process of production of high $k_\bot$ jets, the issue which
will be investigated elsewhere.

\begin{acknowledgments}
We are grateful to the participants of the seminar at BLTP JINR,
Dubna, INP Novosibirsk for critical comments and discussions.

E.~K. and E.~B. acknowledge the support of INTAS grant No. 00366,
RFFI grant No. 03-02-17077 and Grant Program of Plenipotentiary of
Slovak Republic at JINR, grant No. 02-0-1025-98/2005.
\end{acknowledgments}

\newpage

\end{document}